\documentclass[twocolumn,secnumarabic,amssymb, nobibnotes, aps, prb, superscriptaddress]{revtex4-1}

\usepackage{graphicx}
\usepackage{graphics}
\usepackage{xcolor}
\usepackage[normalem]{ulem}
\usepackage{amsmath}

\def\Vbg{V_{\text{G}}}

\def\={\,=\,}

\def\FP{{Fabry-P\'{e}rot }}

\begin{document}

\title{
2$\Phi_{0}$-periodic magnetic interference in ballistic graphene Josephson junctions
}

\author{C. T. Ke} 
\thanks{These authors contributed equally to this work.}
\author{A. W. Draelos}
\thanks{These authors contributed equally to this work.}
\affiliation{Department of Physics, Duke University, Durham, NC 27708, USA.}
\author{A. Seredinski}   
\affiliation{Department of Physics, Duke University, Durham, NC 27708, USA.}
\author{M. T. Wei}
\affiliation{Department of Physics, Duke University, Durham, NC 27708, USA.}
\author{H. Li}
\author{M. Hernandez-Rivera}
\affiliation{Department of Physics and Astronomy, Appalachian State University, Boone, NC 28607, USA.}
\author{K. Watanabe}
\author{T. Taniguchi}
\affiliation{Advanced Materials Laboratory, National Institute for Materials Science, 1-1 Namiki, Tsukuba, 305-0044, Japan.}
\author{M. Yamamoto} 
\affiliation{Center for Emergent Matter Science (CEMS), RIKEN, Wako-shi, Saitama 351-0198, Japan.}
\author{S. Tarucha}
\affiliation{Center for Emergent Matter Science (CEMS), RIKEN, Wako-shi, Saitama 351-0198, Japan.}
\author{Y. Bomze} 
\affiliation{Department of Physics, Duke University, Durham, NC 27708, USA.}
\author{I. V. Borzenets}\email[]{iborzene@cityu.edu.hk}
\affiliation{Department of Physics, City University of Hong Kong, Kowloon, Hong Kong SAR}
\author{F. Amet} 
\affiliation{Department of Physics and Astronomy, Appalachian State University, Boone, NC 28607, USA.}
\author{G. Finkelstein}
\affiliation{Department of Physics, Duke University, Durham, NC 27708, USA.}

\date{\today}%

\begin{abstract}

We investigate supercurrent interference patterns measured as a function of magnetic field in ballistic graphene Josephson junctions. At high doping, the expected $\Phi_{0}$-periodic ``Fraunhofer'' pattern is observed, indicating a uniform current distribution. Close to the Dirac point, we find anomalous interference patterns with an apparent 2$\Phi_{0}$ periodicity, similar to that predicted for topological Andreev bound states carrying a charge of $e$ instead of $2e$. This feature persists with increasing temperature, ruling out a non-sinusoidal current-phase relationship. It also persists in junctions in which sharp vacuum edges are eliminated. Our results indicate that the observed behavior  may originate from an intrinsic property of ballistic graphene Josephson junctions, though the exact mechanism remains unclear.

\end{abstract}

\maketitle

\newpage

Critical current of a Josephson junction subject to a perpendicular magnetic field is known to show decaying oscillations \cite{Tinkham,Dynes1971}. For a uniform supercurrent distribution and a sinusoidal current-phase relation, the pattern of oscillations is identical to that of single-slit Fraunhofer interference. This pattern's periodicity is $\Phi_{0} = h/2e$, the magnetic flux quantum. Measurement of (and deviations from) the ``Fraunhofer pattern" is a conventional way to characterize the uniformity of Josephson junctions, which became particularly relevant for the novel junctions based on 2D materials~\cite{Heersche2007,Miao2007,Du2008,Miao2009}. 

\begin{figure}[h!]
	\center \label{fig1}
	\includegraphics[scale=0.97]{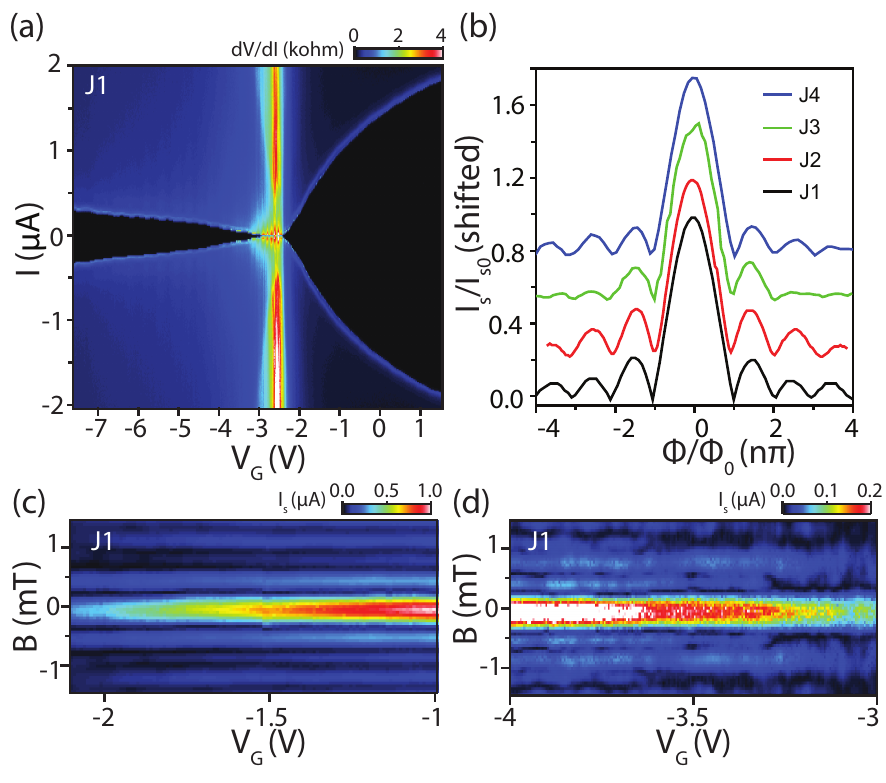}
	\caption{Conventional graphene Josephson junction behavior. (a) Map of differential resistance $dV/dI$ of junction $J_{1}$ as a function of applied current $I$ and back gate voltage $V_{\text{G}}$. The dark region of vanishing resistance near zero bias corresponds to the supercurrent. (b) Regular Fraunhofer interference patterns measured at high density for $J_{1-4}$. 
	(c) Maps of supercurrent $I_\text{S}$ in $J_1$ vs magnetic field $B$ and gate voltage $V_{\text{G}}$, taken at high electron doping. This interference maps demonstrate conventional Fraunhofer patterns with a gate-independent magnetic oscillation period.
	(d) A similar map in the hole-doped regime demonstrating deviations from the Fraunhofer pattern closer to the Dirac point ($V_{\text{D}}=-2.65$ V) and the restored Fraunhofer pattern further away.}
\end{figure}

\begin{figure*}[th!]
	\center \label{fig2}
	\includegraphics[scale=1]{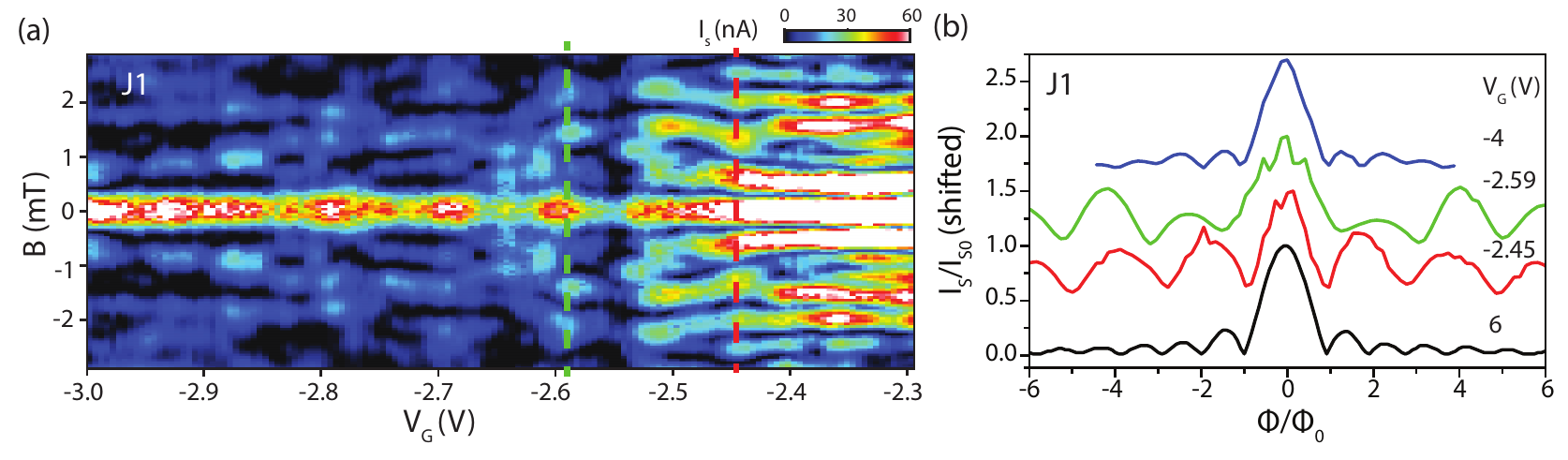}
	\caption{Interference patterns with periodicity doubling. (a) Magnetic interference measurements at low doping for $J_{1}$. Oscillations in the switching current along the gate direction for p-doping result from Fabry-P\'{e}rot resonances in the junction due to PN interfaces near the MoRe contacts. Green and red lines mark cuts displayed in adjacent panel. (b) Line cuts showing normalized $I_\text{S}$ as a function of quantized magnetic flux $\Phi / \Phi_0$. At high electron (black) and hole (blue) doping, a regular oscillation period is observed. However, for certain $\Vbg$ near the Dirac point (red, green) we find a regular pattern of oscillations with a doubled periodic.}
\end{figure*}

Josephson junctions made with topological materials, such as quantum spin Hall insulators, demonstrate marked deviations from the conventional Fraunhofer pattern~\cite{Yacoby2014}. Topological bound states at the superconducting interface are able to support supercurrent along the edges of the sample, resulting in SQUID-like oscillations. Due to the presence of Majorana fermions, these patterns are expected to show a single electron periodicity of $h/e$~\cite{Fu2009,Alicea2012,Zhang}.
However, due to quasiparticle poisoning, this periodicity may not be typically accessible via DC measurements~\cite{Houzet2013, Fu2008, Lee2014}.
Nevertheless, magnetic interference patterns with even-odd modulations have been reported~\cite{YPChen_AB, Kouwenhoven2015, Molenkamp_2016_NatureNano}. Other scenarios are also expected to result in distortions of the interference pattern, including: the presence of a non-uniform supercurrent distribution~\cite{Yacoby2015_DP,Yacoby2015_FP, BenShalom_2017NatComm_edge, deVries2018}, a non-sinusoidal current-phase relation~\cite{Moler_CPR2015,English2016_CPR,Goswami_2017CPR, Harlingen_Natcomm2013}, spin-orbit effects~\cite{Flensberg_2016PRB,Zuo2017,Marcus_2017PRB}, and a non-local supercurrent~\cite{HJLee2015_loc, Barzykin1999, Borghs1998, Ledermann1999, Vandersypen2015, Glazman_2016_edge, Rakyta2016}. Understanding these transport mechanisms may allow one to distinguish trivial 2$\Phi_{0}$-periodic behavior from topological 2$\Phi_{0}$-periodic behavior.

Here, we study ballistic graphene Josephson junctions in several different regimes. We start by measuring the samples at high density, where we find conventional $\Phi_{0}$-periodic Fraunhofer patterns. As the density is lowered, the junctions exhibit a robust lifting of even nodes, resulting in effectively $2\Phi_{0}$-periodic interference patterns. The patterns with even node lifting were repeatably observed in several devices of different geometries, making it highly unlikely that the behavior is a result of an aberrant non-uniform current density. Furthermore, the observed behavior is unaffected by side gates that change the density near the junction/vacuum edges. 
The anomalous patterns persist at elevated temperature where the current-phase relation is expected to be sinusoidal~\cite{English2016_CPR,Goswami_2017CPR}. Note that this system is not expected to host any topological bound states due to its lack of spin-orbit coupling.
Having ruled out the above scenarios, we suggest that this anomalous periodicity may originate from some intrinsic properties of ballistic graphene Josephson junctions. 

We study seven Josephson junctions made of graphene encapsulated in hexagonal boron nitride and contacted by molybdenum-rhenium (MoRe) superconducting electrodes. Fabrication, device, and measurement details are included in the supplementary material. The junctions have different lengths, L, and contact widths, W, as listed in Table S1. Junctions $J_{1-5}$ are conventional rectangular junctions, while $J_{side}$ and $J_{ex}$ include local electrostatic gates along the edges of the junctions. The ballisticity of junctions $J_{1-5}$ was established in depth in a previous study~\cite{Borzenets2016}. For the first part of this paper we focus on $J_{1}$ for clarity and brevity. 

The differential resistance of $J_{1}$ as a function of applied DC bias current $I$ and back gate voltage $\Vbg$ is shown in Fig. 1a. The black region roughly symmetric about $I=0$ corresponds to the superconducting state, in which the junction resistance vanishes. The transition from the superconducting state to the normal state occurs at the switching current, $I_\text{S}$. At high carrier density, the switching current increases proportionally with the number of conducting modes; it reduces to its minimum around the Dirac point ($V_D\approx-2.65$V).
Conventional Fraunhofer patterns are observed at high electron density in all junctions (Fig. 1b). In $J_1$, this regime persists for densities $n \gtrsim 5 \times 10^{10}$ cm$^{-2}$ ($V_{\text{G}} - V_{\text{D}} \gtrsim 0.7$ V). This is shown in Fig. 1c, which demonstrates that the oscillations remain unchanged for most of the $V_\text{G}$ range. In the case of hole doping, PN junctions are formed in the graphene due to local N-doping by the MoRe contacts. This results in \FP oscillations, which manifest as resonances in $I_\text{S}$ measured versus $\Vbg$~\cite{Vandersypen2015, Geim2015_QHSC} (see Supplementary Information). In this regime, anomalies in the Fraunhofer pattern are clear, as seen in the contrast between Fig. 1c and 1d. Previous work has attributed these anomalous patterns to different \FP resonances in bulk and edge modes, resulting in edge-dominated (SQUID-like) interference when the bulk transmission is low~\cite{Yacoby2015_FP}.

In this paper, we explore the regime of very small densities, $n \lesssim 2.5 \times 10^{10}$ cm$^{-2}$, presented in Fig. 2a. Here, we find that 2$\Phi_0$-periodic interference patterns arise as shown in Fig. 2b, which correspond to the vertical cross-sections of Fig. 2a at $\Vbg=-2.45$ V and $-2.59$ V. In these two curves (red and green), all even nodes are completely lifted and the overall periodicity changes from $0.35$ mT to $0.68$ mT as compared to results at high electron and hole density (blue and black). The interference patterns retain this regular 2$\Phi_0$ period for several flux quanta and thus the pattern change cannot be explained by a randomly distorted current distribution. The regions of periodicity doubling also persist for a significant range of $\Vbg$ (Fig. 2a), further indicating that disordered current distribution is not likely to be the cause.

We would like to point out that as the result of the pattern change, the width of the side lobes becomes roughly equal to the width of the central peak, whose width stays constant. While a central maximum with the same width as the side maxima indicates a SQUID-like interference pattern, the observed behavior would correspond to a SQUID with a 4$\pi$ periodic current-phase relation (CPR). However, previous CPR studies on ballistic graphene Josephson junctions show a 2$\pi$-periodic CPR persisting through the Dirac point~\cite{English2016_CPR, Goswami_2017CPR}. Alternatively, 2$\Phi_0$-periodic interference pattern could be a result of crossed Andreev reflections, in which the electron and the Andreev-reflected hole propagate along the opposite sides of the junction, as discussed Ref.~\cite{deVries2018} for topologically trivial InAs. We explore (and refute) possible contributions from the edges in our samples the following.

\begin{figure}[ht]
		\center \label{fig4}
		\includegraphics[scale=1]{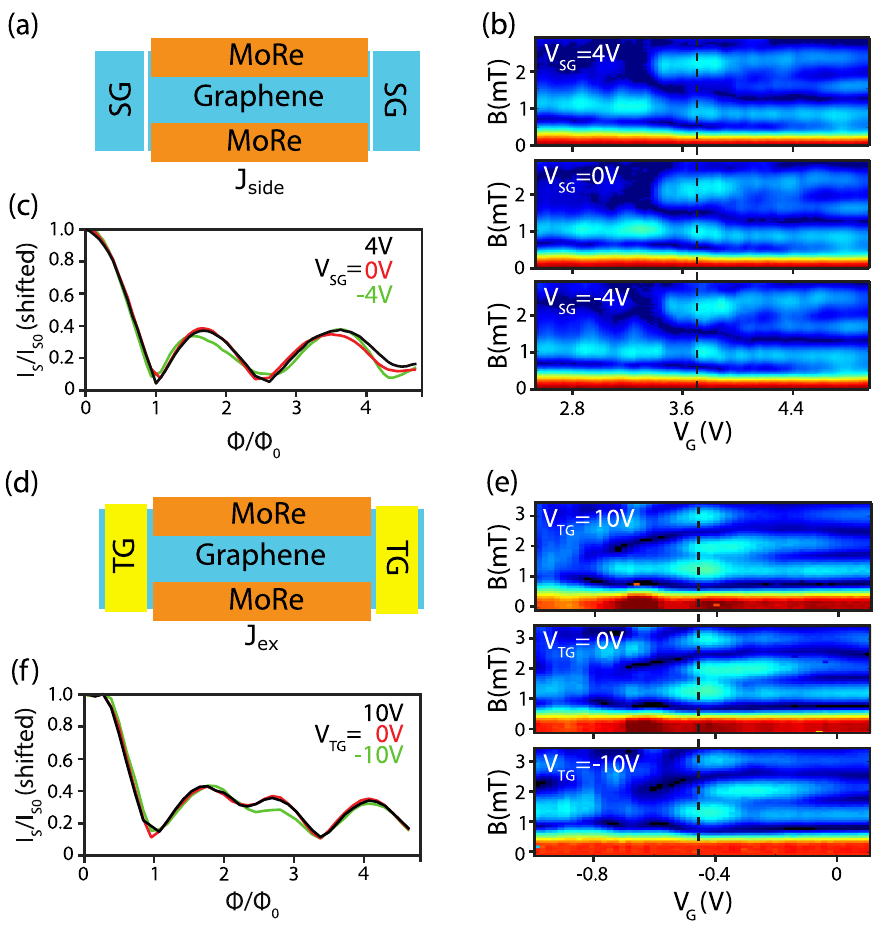}
		\caption{Interference in junctions with edge density control. (a) Schematic of $J_{side}$ showing two side gates which modify the potential at the vacuum edges.
		(b) The normalized supercurrent as a function of magnetic field and gate voltages in $J_{side}$. The density on both edges is increased (decreased) by $\sim 4\times 10^{11}$ cm$^{-2}$ for $V_{SG}=4$ V ($-4$ V). Little influence on the interference pattern is seen. (c) Line cuts from the dashed lines in (b) showing lifting of the even nodes. 
		(d) Schematic of $J_{ex}$ with extended areas on both sides of the mesa and top gates to control the local density. 
		(e) As (b) but for $J_{ex}$, applying identical voltages to both local top gates. Here, the local density is tuned dramatically by $\sim 5 \times 10^{12}$ cm$^{-2}$ for $V_{TG}=\pm 10$ V. Again, the influence on the interference pattern is minor. 
		(f) Line cuts from the dashed line in (e). Again, there is no influence on the anomalous period by the local gates. Hence we conclude that the anomalous periodicity is not caused by trivial edge channels.}
	\end{figure}

\begin{figure}[t]
	\center \label{fig3}
	\includegraphics[scale=0.95]{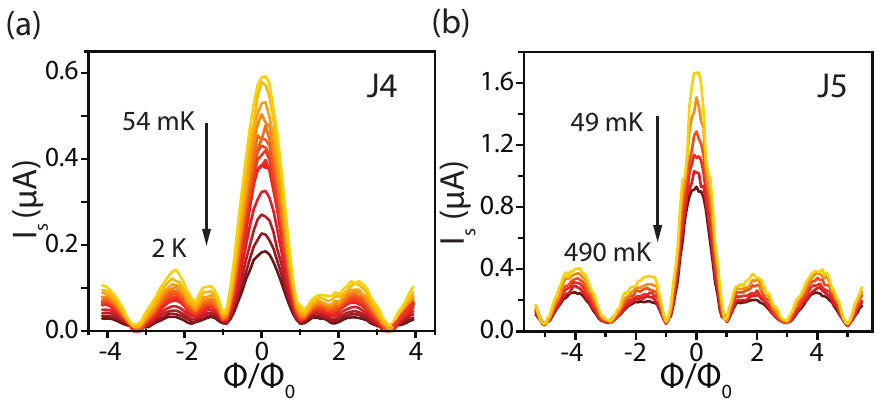}
	\caption{Temperature dependence of the Fraunhofer pattern in (a) $J_4$ and (b) $J_5$. Both patterns show that only the amplitude of the switching current is changing with increasing temperature, while the anomalous pattern remains qualitatively unchanged. This observation indicates that the phenomenon is not attributable to an anomalous current-phase relation, which would have been suppressed at elevated temperatures.}
\end{figure}

Previous studies have demonstrated a significant density buildup along the vacuum edges of graphene devices~\cite{Silvestrov2008, Vera-Marun2013, Cui2016}.
In order to determine whether edge effects are responsible for the anomalous interference patterns, we employ two types of local gates which directly affect the edge carrier density. In $J_{side}$, self-aligned side gates are made from the same graphene crystal as the Josephson junction by etching a narrow gap ($\lesssim 100$ nm) between the two (Fig. 3a)~\cite{Seredinski2019}. As the side gates are very close to the edge of the junction, they are highly efficient, allowing us to change the local density by approximately $\pm 4\times 10^{11} {\text{cm}^{-2}}$ relative to the bulk. In $J_{ex}$, the graphene mesa extends several microns beyond the junction on both edges. The carrier densities of these extended regions are controlled by two local top-gates, which come to within $\sim 100$ nm of the contacts and tune the local density by $\pm 5 \times 10^{12}$ cm$^{-2}$ (Fig. 3b). This sample geometry eliminates contributions from the trivial edge states introduced by the termination of the graphene lattice, as the physical edges of the graphene are several microns long -- greater than the superconducting coherence length -- and therefore do not contribute to the supercurrent.

Both $J_{side}$ and $J_{ex}$ show lifting of even nodes near the bulk Dirac point, while the first and third nodes remain pronounced. Interestingly, the first node also remains pinned at a fixed value of $\Phi=\Phi_0$ while the rest of the pattern is strongly distorted. This observed anomalous periodicity is robust and persists over the full range of side or top gate voltages. In fact, the interference maps as a function of gate voltage and magnetic field (Fig. 3c,d) remain roughly unchanged as the top or side gates are applied, except for a small shift along the back gate axis, induced by their overall electrostatic influence. 

In particular, the anomalous pattern persists in $J_{side}$ (Fig. 3c,e) as the side gate voltages are applied in both the positive and negative directions, such that the edges of the junction acquire a carrier density ($\sim \pm 10^{11} {\text{cm}^{-2}}$) that greatly exceeds the density in the bulk ($\sim 10^{10}{\text{cm}^{-2}}$). $J_{ex}$, which certainly does not have supercurrent mediated by trivial states at the graphene-vacuum edge, demonstrates even node lifting through zero top gate voltage, when density at the edges is close to the density in the bulk (Fig. 3e,f). The lack of side gate voltage sensitivity appears to rule out the contribution of the edge as the cause of the anomalous interference pattern. In the following we consider alternative mechanisms that are known to modify the magnetic interference patterns and discuss if they could explain the observed behaviour.

Geometric effects resulting in non-local supercurrent may yield unconventional interference patterns. For high aspect ratio junctions ($L/W\gtrsim1$), supercurrent from trajectories with reflections from the vacuum edges of the junction cannot be neglected.  This non-local supercurrent distribution can yield a magnetic interference pattern with a 2$\Phi_{0}$ periodicity~\cite{Barzykin1999,Vandersypen2015,Glazman_2016_edge}. 
However, all of our devices have small geometric ratios $0.07<L/W<0.33$ (see Table S1 in the Supplementary Information). Furthermore, the results from this non-local distribution would change the period of interference, but would not yield a pattern with a central maximum of the same width as the side maxima. Therefore, this cannot explain the pattern with even node lifting that we observe.

Any possible explanations involving a nonsinusoidal CPR can be ruled out by measurements at increased temperature, where a $2\pi$-periodic sinusoidal relation should be recovered~\cite{English2016_CPR,Goswami_2017CPR}. Such temperature dependence is presented in Fig. 4 for junctions $J_4$ and $J_5$ showing the persistence of anomalous periodicity up to 2 K. The variety of junction lengths studied provides further evidence that this behavior is not related to CPR. Indeed, our devices range from the short ($L < \xi_0$, where $\xi_0=\frac{\hbar v_F}{\Delta}$ is the superconducting coherence length) to the long ($L > \xi_0$) ballistic regime. Junctions in these opposing limits are expected to display different CPR, suggesting that the consistent behavior across all of our devices cannot originate with the CPR.

A disordered supercurrent density concentrated at several locations along the junction could explain the lifting of the nodes of the interference pattern. Indeed, the current density becomes less uniform close to charge neutrality. However, it is unlikely that such disorder would result in preferential lifting of the even nodes.

We are led to conclude that the anomalous pattern may arise from some intrinsic property of graphene. Indeed, carrier trajectories in graphene can be influenced by the valley degree of freedom, and by specular Andreev reflections \cite{Beenakker2006}. 
It is not clear whether either could produce the observed lifting of the even nodes and the resulting quasi-$2\Phi_0$-periodic patterns. 

In summary, we have explored magnetic interference patterns throughout different density regimes for ballistic graphene Josephson junctions. At high carrier density, a regular Fraunhofer pattern is observed, indicating a uniform current distribution across the width of the junctions. Remarkably, at lower carrier densities we observe a robust lifting of even nodes in the interference patterns of all devices. Temperature dependence, different junction lengths, and control of trivial edge channels were exhaustively considered; in all cases regions of anomalous periodicity persisted. Our observations rule out likely explanations such as a non-sinusoidal CPR, large current densities at the graphene-vacuum edges, and non-local supercurrent contributions. While a topological state with a 4$\pi$ periodic CPR would explain our observation, no such state has been predicted to exist in graphene. Further studies of the observed behavior are clearly needed. 

\begin{acknowledgements}
Low-temperature electronic measurements performed by C.T.K., A.W.D. and G.F. were supported by the Office of Basic Energy Sciences, U.S. Department of Energy, under Award DE-SC0002765. Lithographic fabrication and characterization of the samples performed by A.S. and M.T.W. were supported by ARO Award W911NF16-1- 0122 and NSF awards ECCS-1610213 and DMR-1743907. H. Li, M.H.R., and F.A. acknowledge the ARO under Award W911NF-16-1-0132. A.W.D. was supported by the NSF graduate research fellowship DGF1106401. K.W. and T.T. acknowledge the Elemental Strategy Initiative conducted by the MEXT, Japan and the CREST (JPMJCR15F3), JST. S.T. and M. Y. acknowledges KAKENHI (GrantNo. 38000131, 17H01138). I.V.B. acknowledges CityU New Research Initiatives/Infrastructure Support from Central (APRC): 9610395, and the Hong Kong Research Grants Council (ECS) Project: 9048125. This work was performed in part at the Duke University Shared Materials Instrumentation Facility (SMIF), a member of the North Carolina Research Triangle Nanotechnology Network (RTNN), which is supported by the National Science Foundation (Grant ECCS-1542015) as part of the National Nanotechnology Coordinated Infrastructure (NNCI). 
\end{acknowledgements}

\pagebreak
\clearpage
\onecolumngrid
\begin{center}
\textbf{\huge Supplementary Information: Robust anomalous magnetic interference in graphene Josephson junctions}

\end{center}

\newcommand{\beginsup}{%
        \setcounter{equation}{0}
        \renewcommand{\theequation}{S\arabic{equation}}%
        \setcounter{table}{0}
        \renewcommand{\thetable}{S\arabic{table}}%
        \setcounter{figure}{0}
        \renewcommand{\thefigure}{S\arabic{figure}}%
     }

\beginsup


	
	
\twocolumngrid
\section{Devices under study}

We fabricated Josephson junctions using exfoliated monolayer graphene crystals encapsulated in hexagonal boron nitride. The junctions’ contacts are made by sputtering molybdenum-rhenium (MoRe), a type-II superconductor. Further details of sample fabrication are described in our previous work~\cite{Sup_Amet_QHSC}. All samples are measured in a cryogen-free dilution refrigerator with a base temperature of 35 mK. To determine the supercurrent, a bias current is continuously, linearly ramped through the junction at a repetition rate of $\sim$100 Hz and the voltage difference is measured across the contacts, yielding I-V curves. The ramp step size has a resolution of 0.3 nA, and it has been confirmed that the filtering/wiring of the device does not significantly disturb the ramp shape. The switching current $I_\text{S}$ is then extracted as a function of back gate voltage $V_\text{G}$ and magnetic field $B$. Due to the possibility of flux trapping in the magnet or sample~\cite{ieee}, the magnetic field is limited to $\pm$ 5 mT.

\begin{table}[h!]
	\begin{center}
		\caption{List of samples}
		\label{tab:table1}
		\begin{tabular}{|c|c|c|c|}
			\hline
			Device name & Length & Width & Ratio: L/W \\ 
			\hline
		    $J_{1}$  & 0.65$\,$$\mu m$ & 4.5$\,$$\mu m$ &  0.144  \\ 
            $J_{2}$  & 0.3$\,$$\mu m$ & 2.4$\,$$\mu m$ & 0.125  \\
            $J_{3}$  & 0.2$\,$$\mu m$ & 3$\,$$\mu m$ & 0.067  \\
			$J_{4}$  & 0.4$\,$$\mu m$ & 3$\,$$\mu m$ & 0.133  \\
			$J_{5}$  & 1$\,$$\mu m$ & 3$\,$$\mu m$ & 0.333  \\
			$J_{side}$ & 0.5$\,$$\mu m$ & 3$\,$$\mu m$ & 0.166    \\
			$J_{ex}$ & 0.5$\,$$\mu m$ & 3$\,$$\mu m$ & 0.166    \\
			\hline
		\end{tabular}
	\end{center}
\end{table}

\section{Fabry-P\'{e}rot regime}

The normal resistance $R_\text{N}$ and switching current $I_\text{S}$ of $J_1$ are plotted in Figure S1 as a function of gate voltage, showing a sharp resistance peak indicating the location of the Dirac point. Oscillations in both the normal resistance and switching current are visible. These oscillations result from the work function mismatch between the MoRe superconducting leads and the graphene, which yields PN interfaces: MoRe locally n-dopes the graphene; and (in a ballistic device) a Fabry-P\'{e}rot cavity develops when the graphene bulk is p-doped by the backgate $V_\text{G}$ \cite{Sup_Vandersypen2015}. Figure S2 plots oscillations in the normal conductance $\Delta G_\text{N}$ and the switching current $\Delta I_\text{S}$ versus the gate voltage $V_\text{G}$ for device $J_1$, computed by subtracting a linear fit (fitted in the region displayed) from the measured $G_\text{N}$ ($I_\text{S}$) (Figure S1). This procedure isolates the oscillations, and one clearly observes regular resonances in both the supercurrent and the normal conductance. $G_\text{N}$ and $I_\text{S}$ oscillate in-phase with each another, as expected~\cite{Sup_Geim2015_QHSC}.
\begin{figure}[h!]
		\center \label{figS1}
		\includegraphics[scale=0.95]{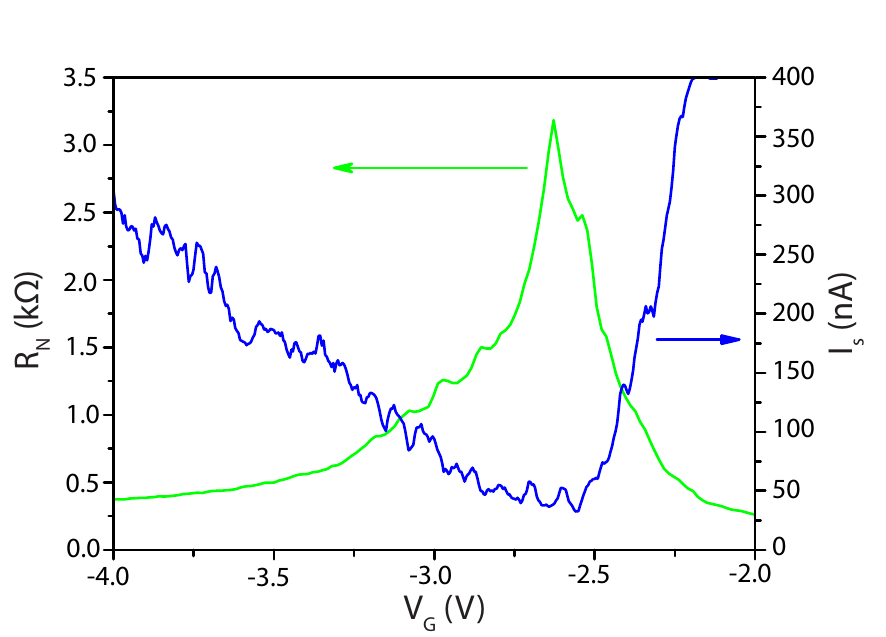}
		\caption{The normal resistance $R_\text{N}$ (green) and switching current $I_\text{S}$ (blue) as a function of the gate voltage $V_\text{G}$ for device $J_1$. The sharp peak of $R_\text{N}$ indicates the DP at $V_{\text{G}} =-2.65$ V. Both resistance and critical current show oscillation  behavior in the \FP regime from -2.65 V to -3.0 V.  }
	\end{figure}

\begin{figure}[h!]
		\center \label{figS2}
		\includegraphics[scale=0.95]{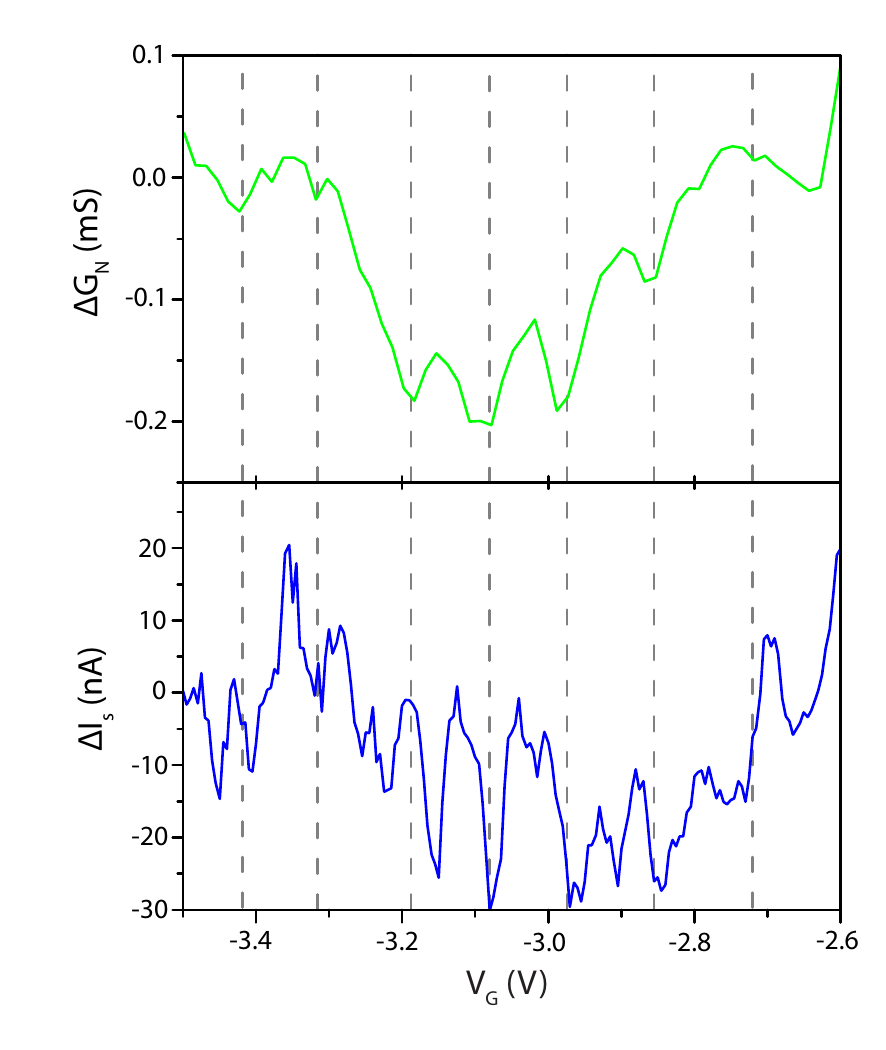}
		\caption{Local change in normal conductance $\Delta G_\text{N}=1/\Delta R_\text{N}$ and switching current $\Delta I_\text{S}$ versus gate voltage $V_\text{G}$ showing \FP oscillations for device $J_1$. $\Delta G_\text{N}$ and $\Delta I_\text{S}$ were computed by subtracting a linear fit from the measured data in order to amplify the visibility of oscillations. As expected, $\Delta I_\text{S}$ oscillates in phase with $\Delta G_\text{N}$.}
	\end{figure}

\section{Magnetic Interference Pattern at the Dirac point.}
Figure S3 shows the magnetic interference pattern of junction J1 taken at the Dirac point ($V_{\text{G}} \approx-2.64$ V). The dependence of critical current $I_\text{S}$ on $\Phi/\Phi_0$ is quite complex here. The critical current is never fully suppressed until after the third side-lobe. This is indicative of a non-uniform current density distribution~\cite{Sup_Tinkham}, which is consistent with the fact that very few conducting channels are available at the Dirac point. Conduction at the Dirac point may be affected by local impurity doping and is therefore expected to be non-uniform~\cite{sup_puddles}. 
Note that the side lobe peak locations reveal an increase in period similar to that discussed in the main text. However, the period is not precisely doubled, but rather multiplied by $\sim 1.5$. It is possible that the same effect causing the doubling discussed in the main text is at play here as well, but complicated by the highly disordered pattern making analysis difficult. Note that due to this change in period, the pattern observed is not a simple SQUID-like pattern as one would expect for pure edge transport~\cite{sup_Yacoby2015_FP}.

\begin{figure}[h!]
		\center \label{figS3}
		\includegraphics[scale=1]{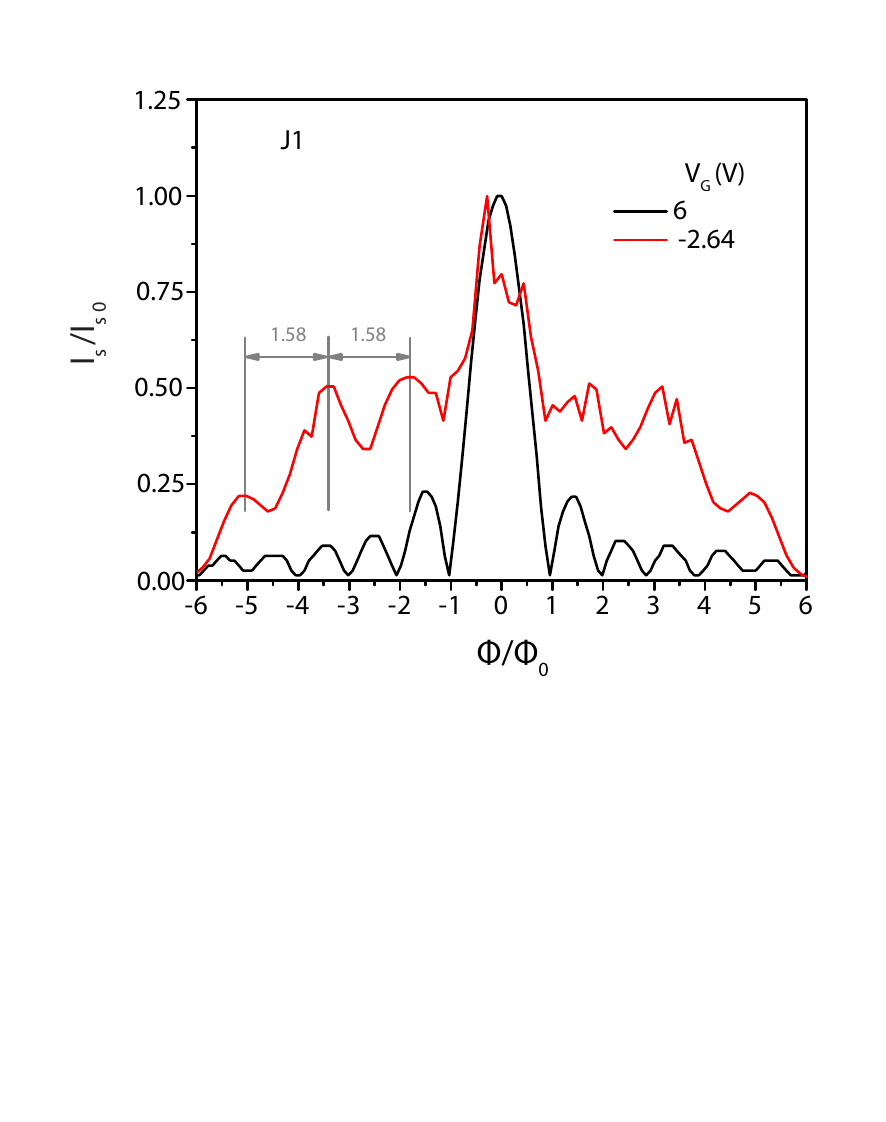}
		\caption{The interference patterns at high doping regime (black) and Dirac point (red). The switching current axis is normalized by the zero field value $I_{S0}=I_\text{S}(B=0)$. While the high doping regime demonstrates a typical Fraunhoffer pattern, at the Dirac point the pattern is significantly distorted. The overall lifting of the nodes can be attributed to a disordered current density distribution. A change in periodicity is also observed by about a factor of $\sim 1.5$.}
	\end{figure}

\section{Magnetic interference for other junctions}
Here we show the magnetic interference patterns for devices $J_2$ and $J_3$. Following the format of Figure 2 in the main text, Figure S4 shows the interference patterns for $J_2$:(a),(c) and $J_3$:(b),(d). The gate maps focus on the regions around the Dirac point, and the red and green lines indicate gate voltage $V_\text{G}$ regions of periodicity doubling. While not as clean as the devices shown in the main text, $J_2$ and $J_3$ also show clear lifting of even nodes.

	\begin{figure}[h!]
		\center \label{figS4}
		\includegraphics[scale=0.8]{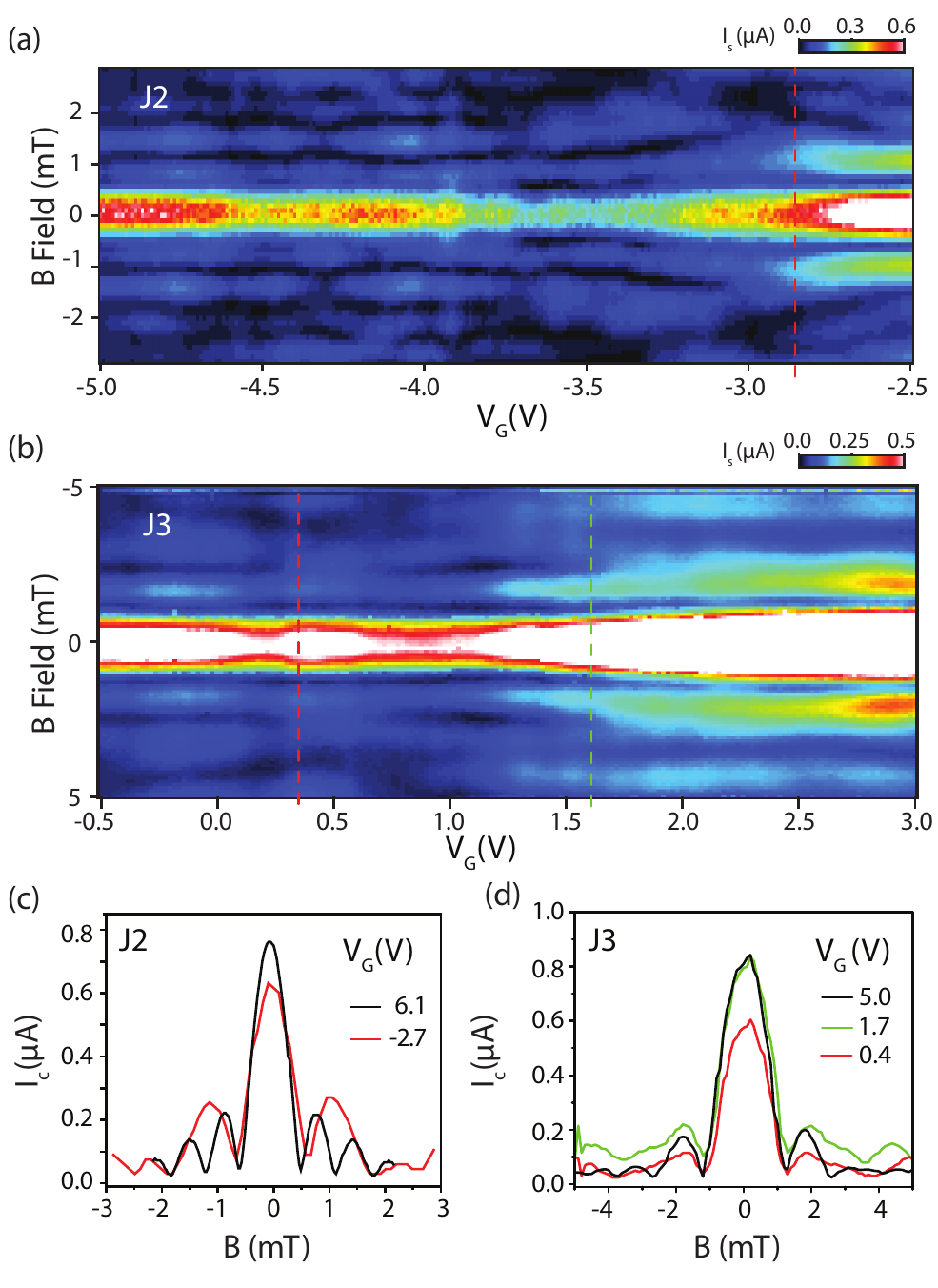}
		\caption{Interference patterns with periodicity doubling. (a) and (b): Magnetic interference measurements at low doping for $J_{2}$ and $J_{3}$. Green and red lines mark cuts displayed in lower panels. (c) and (d): Line cuts showing supercurrent $I_\text{S}$ as a function of quantized magnetic field $B$ for junctions  $J_{2}$ and $J_{3}$, respectively.  At high electron doping, a regular oscillation period is observed (black line). However, for certain $\Vbg$ near the Dirac point (red, green) we find a robust lifting of the second node, resulting in an effective 2$\Phi_0$ periodicity.}
	\end{figure}

\pagebreak

\end{document}